\documentclass[aps,pra,twocolumn,groupedaddress,floatfix,showpacs]{revtex4-1}
\usepackage{graphicx}
\usepackage{amsmath}
\usepackage{color}
\def\ket#1{|#1\rangle}
\def\bra#1{\langle#1|}

\begin{document}
\title{Energy-Resolved Information Scrambling in Energy-Space Lattices}

\author{S. Pegahan, I. Arakelyan and J. E. Thomas}

\affiliation{$^{1}$Department of  Physics, North Carolina State University, Raleigh, NC 27695, USA}

\date{\today}

\begin{abstract}
Weakly interacting Fermi gases simulate spin-lattices in energy-space, offering a rich platform for investigating information spreading and spin coherence in a large many-body quantum system. We show that the collective spin vector can be determined as a function of energy from the measured spin density, enabling general energy-space resolved protocols. We measure an out-of-time-order correlation function in this system and observe the energy dependence of the many-body coherence.
\end{abstract}

\maketitle

Trapped, weakly interacting Fermi gases provide a new paradigm for the study of many-body physics in a large quantum system containing $N\simeq 10^5$ atoms with a tunable, reversible Hamiltonian~\cite{DuSpinSeg2,ThywissenDynamicalPhases}. In this system, coherent superpositions of two hyperfine states behave as pseudo-spins and the s-wave scattering length is magnetically tuned to nearly vanish~\cite{DuSpinSeg1,DuSpinSeg2,SaeedPRASpinECorrel}. The corresponding collision rate is negligible, so that single atom energies are conserved~\cite{DuSpinSeg2,Piechon,MuellerWeaklyInt,LaloeSpinReph} over the experimental time scale. The conserved single particle energy states label the ``sites" of an effective energy-space lattice, simulating a variety of spin-lattice models~\cite{KollerReySpinDep}. Interactions are effectively long range in energy-space~\cite{LewensteinDynLongRange,KollerReySpinDep,SaeedPRASpinECorrel},  important for new studies of information scrambling in a far from equilibrium, nearly zero temperature regime~\cite{ReyNatPhys2017} and for applications to fast scrambling~\cite{FastScramblingColdAtoms} and ``out-of-equilibrium" dynamics in spin-lattice systems~\cite{OutofEq}. However, measurements in weakly interacting Fermi gases~\cite{DuSpinSeg1,DuSpinSeg2,SaeedPRASpinECorrel,Piechon,MuellerWeaklyInt,LaloeSpinReph,ThywissenDynamicalPhases} have been limited to the spatial profiles of the collective spin density or the total number of atoms in each spin state, precluding observation of many-body correlations in chosen sectors of the energy-space lattice.

Of particular interest is the measurement of out-of-time-order correlation (OTOC) functions in  weakly interacting Fermi gases. Certain OTOC functions~\cite{Schleier-Smith2017,SwingleHighlight,JLiNMR,ReySlowFastScrambling} can serve as entanglement witnesses and to quantify coherence and information scrambling in quantum many-body systems~\cite{ReyNatPhys2017,ReyPRL2018}. Originally, OTOC measurements were performed by reversing the time evolution of the many-body state in nuclear magnetic resonance experiments at high temperatures, where the initial state is described by a density operator and high order quantum coherence was observed~\cite{Pines1985MultCoh}.  New OTOC studies have been done in trapped ion systems containing relatively small numbers of atoms, where the individual sites are nearly equivalent, and the initial state is pure~\cite{ReyNatPhys2017}. Related methods have been developed for systems containing up to 100 atoms~\cite{ReyNat2019}, but the application of OTOC measurement to trapped ultracold gases has remained a challenge.

In this Letter, we report the demonstration of a general method for performing energy-resolved measurements of the collective spin vector in a harmonically-trapped weakly-interacting Fermi gas. We show that OTOC measurements can be implemented in this system and we extract many-body coherence in energy-resolved sectors, paving the way for new protocols, such as time-dependent energy-space correlation measurements.

\begin{figure*}[htb]
\begin{center}\
\hspace*{-0.0in}\includegraphics[width=6.5in, trim = {0 1 0 0}, clip]{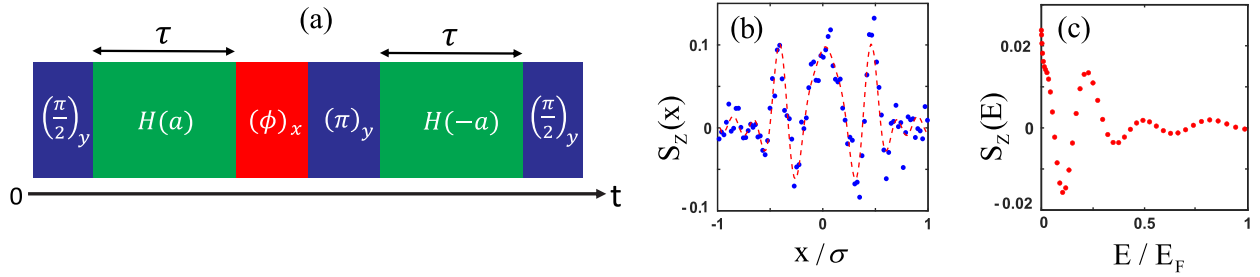}
\end{center}
\caption{Energy-resolved out-of-time-order correlation (OTOC) measurement. The system is initially prepared in a pure state, with the spins for atoms of energy $E_1,E_2,...E_N$ polarized along the $-z$ axis; (a) OTOC sequence, after which the spatial profiles of the $\uparrow_z$ and $\downarrow_z$ states are measured for each cloud by resonant absorption imaging; (b) ``single-shot" spin density profile $S_z(x)$ (blue dots). For this measurement, the scattering length in the Hamiltonian $H(a)$ is $a=4.24\,a_0$,  $\phi =\pi$, and $\sigma=345\,\mu$m. \\ (c) An inverse-Abel transform of the spatial profile (blue dots) extracts the single-shot energy-resolved spin density $S_z(E)$ (red dots). An Abel transform of $S_z(E)$ yields the red-dashed curve shown in (b), consistent with the data.
\label{fig:protocol}}
\end{figure*}

In the experiments~\cite{Supplement}, we begin with a degenerate cloud of $^6$Li containing a total of $N=6.5\times 10^4$ atoms in a single spin state. The cloud is confined in a harmonic, cigar-shaped optical trap, with oscillation frequencies $\omega_x/2\pi=23$ Hz along the cigar x-axis and $\omega_r/2\pi=625$ Hz in the transverse ($y,z$) directions. The corresponding Fermi temperature $T_F=0.73\,\mu$K and $T/T_F=0.32$.

We employ the two lowest hyperfine-Zeeman states, which are denoted by $\ket{1}\equiv\ket{\!\uparrow_z}$ and $\ket{2}\equiv\ket{\!\downarrow_{z}}$. The cloud is initially prepared in state $\ket{\!\downarrow_{z}}$ in a bias magnetic field of $528.53$ G, where the s-wave scattering length $a_{12}\equiv a=4.24\,a_0$~\cite{SaeedPRASpinECorrel}.  In this case, the largest possible collision rate $\gamma_c$ in the Fermi gas arises for an incoherent mixture with $N/2$ atoms in each of two spin states. We find $\gamma_c<1.7\times 10^{-3}\,{\rm s}^{-1}$~\cite{GehmCollisionRate}, which is negligible for the experimental time scale $<1\,$s. Hence, the single particle energies are conserved and the energy distribution is time independent,  as observed in the experiments~\cite{SaeedPRASpinECorrel,Supplement}.

The Hamiltonian for the confined weakly interacting Fermi gas can be approximated as a one-dimensional (1D) spin ``lattice" in energy space~\cite{SaeedPRASpinECorrel},
\begin{equation}
H(a)=a\!\sum_{i,j\neq i}g_{ij}\,{\mathbf s}_i\cdot{\mathbf s}_j-\sum_{i}\Omega_{i}\,s_{zi}
\label{eq:1.1}
\end{equation}
where we take $\hbar\equiv 1$. We associate a ``site" $i$ with the energy $E_i\!=\!(n_i\!+\!1/2)\,h\nu_x$ of an atom in the i$^{\rm th}$ harmonic oscillator state along the cigar axis $x$. For each $E_i$, we define a dimensionless collective spin vector $\mathbf{s}_i=\sum_{\alpha_i}\mathbf{s}_{\alpha_i}$, where the sum over $\alpha_i$ includes the occupied transverse ($n_y,n_z$) states for fixed $n_i$. As $k_B T_F/\hbar\omega_x\simeq 650$, the average number of atoms at each site is $N/650\simeq 100$~\cite{NumberperSite}.

The first term in Eq.~\ref{eq:1.1} is a site-to-site interaction, proportional to the s-wave scattering length $a$ and to the overlap of the harmonic oscillator probability densities for colliding atoms,  $g_{ij}\propto \int dx\,|\phi_{E_i}(x)|^2\,|\phi_{E_j}(x)|^2\propto 1/\sqrt{|E_i-E_j|}$, which is an effective long range interaction in the energy lattice~\cite{SaeedPRASpinECorrel}. For a zero temperature Fermi gas, the average interaction energy is $a\bar{g}=3.8\,\Omega_{MF}$~\cite{gAv}, where the mean field frequency~\cite{SaeedPRASpinECorrel} for our experimental parameters is $\Omega_{MF}/2\pi\simeq 0.5$ Hz, i.e., $a\bar{g}/2\pi\simeq 1.9$ Hz.

The second term in Eq.~\ref{eq:1.1} is an effective site-dependent Zeeman energy, arising from the quadratic spatial variation of the bias magnetic field along $x$, which produces a spin-dependent harmonic potential. As $\omega_r/\omega_x=27$,  the corresponding effect on the transverse ($y,z$) motion is negligible, so that all atoms at site $i$ have the same Zeeman energy. In Eq.~\ref{eq:1.1}, $\Omega(E_i)\equiv\Omega_{i}=\Omega'\,E_i+\Delta'$, where $\Omega'=-\delta\omega_x/\hbar\omega_x$, with $\delta\omega_x/2\pi=14.9$ mHz for our trap~\cite{SaeedPRASpinECorrel}. For atoms with the mean energy $\bar{E}_x\simeq k_B T_F/4$, $\Omega'\,\bar{E}_x/2\pi\simeq 2$ Hz.  We define $\Delta'\equiv\Delta-\Omega'\,\bar{E}_x$, where $\Delta$ is the global detuning and $\Delta = 0$ corresponds to $\Omega_i=0$  for the mean energy, $E_i=\bar{E}_x$.

A key feature of our experiments is the extraction of energy-resolved spin densities $n_{\uparrow_z,\downarrow_z}(E)$ by inverse Abel-transformation of the corresponding 1D spatial profiles $n_{\uparrow_z,\downarrow_z}(x)$,  which are obtained from absorption images of a single cloud. The transform method requires a continuum approximation, which is justified for the x-direction, where $k_BT_F/\hbar\omega_x=650$. Further, we require negligible energy space coherence, i.e., the atomic spins remain effectively localized in their individual energy sites.  This assumption is justified by the very small transition matrix elements $< 10^{-4}\,\hbar\omega_x$~\cite{MatrixEl} between three dimensional harmonic oscillator states, which arise from short range interactions between two atoms~\cite{Supplement}.

In this regime, the spatial profile for each spin state $n_\sigma(x)$, $\sigma\equiv\,\uparrow_z,\downarrow_z$, is an Abel transform of the corresponding energy profile $n_\sigma(E)$~\cite{Supplement},
\begin{eqnarray}
n_\sigma(x)&=&\int\! dE\,|\phi_E(x)|^2\,n_\sigma(E)\nonumber\\
&=&\frac{\omega_x}{\pi}\int_0^\infty\!\!\! dp_x\,n_\sigma\!\left(\frac{p_x^2}{2m}+\frac{m\omega_x^2}{2}x^2\right).
\label{eq:spinupdowndensity}
\end{eqnarray}
In Eq.~\ref{eq:spinupdowndensity}, the last form is obtained by using a WKB approximation for the harmonic oscillator states $\phi_E(x)$~\cite{Supplement}. An inverse Abel-transform~\cite{Supplement,NewAbelInversion} of $n_\sigma(x)$ then determines $n_\sigma(E)$ with a resolution $\Delta E\simeq 0.04\,E_F$~\cite{Supplement}.

For the protocol of Fig.~\ref{fig:protocol}(a), discussed in detail below, Fig.~\ref{fig:protocol}(b) shows the  measured {\it single-shot} spin density,  $S_z(x,\phi)=[n_{\uparrow_z}(x,\phi)-n_{\downarrow_z}(x,\phi)]/2$, in units of the central total spin density $n(0)$.   Fig.~\ref{fig:protocol}(c) shows the corresponding single-shot $S_z(E,\phi)$, obtained by inverse-Abel transformation of $S_z(x,\phi)$. We see that  $S_z(E,\phi)$ appears smooth compared to the single-shot spin density $S_z(x,\phi)$, which requires averaging over several shots to obtain a smooth profile. To check that the inverse-Abel transform has adequate energy resolution, we Abel transform the extracted $S_z(E,\phi)$, yielding the red-dotted curve of Fig.~\ref{fig:protocol}(b), which is consistent with the measured density profile~\cite{Supplement}.

\begin{figure*}[htb]
\begin{center}\
\hspace*{-0.25in}\includegraphics[width=6.0in,trim = {0 1 0 0}, clip]{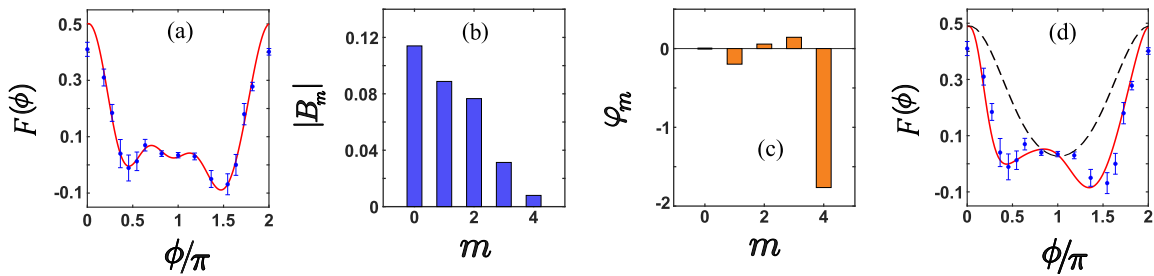}
\end{center}
\caption{Total collective spin projection $S_z$  versus rotation angle $\phi$ without energy restriction. (a)  $F(\phi)=\frac{1}{2}(N_{\uparrow_z}-N_{\downarrow_z})/(N_{\uparrow_z}+N_{\downarrow_z})$ (blue dots) for a measured scattering length $a_{\rm meas}=4.24\,a_0$. The red solid curve is the fit of Eq.~\ref{eq:18.6}, which determines the magnitudes of the coherence coefficients $|B_m|$ (b) and corresponding phases $\varphi_m$ (c); (d) Fit of the mean field model of Ref.~\cite{SaeedPRASpinECorrel} to the data (blue dots) for a global detuning $\Delta = 0$ with $a=a_{\rm meas}$  (black-dashed) and with $a=2.63\,a_{\rm meas}$ (red-solid).
\label{fig:TotalSz}}
\end{figure*}

Our experimental OTOC protocol, Fig.~\ref{fig:protocol}(a), applies a rotation $\phi$ to the total interacting spin system in between forward and time-reversed evolutions. Then, a measurement of $s_{zi}$ is performed to diagnose the effects of the rotation on the spins at ``site i" in energy space.  We start by preparing a fully z-polarized state $\ket{\!\!\downarrow_{z1}\downarrow_{z2}...\downarrow_{zN}}\equiv\ket{\psi_{z0}}$ in a bias magnetic field $B_1=528.53$ G, where the scattering length $a_1\equiv a=4.24\,a_0$. Then we apply a $0.5$ ms radio-frequency $(\pi/2)_y$ pulse (defined to be about the y-axis), which is resonant with the \mbox{$\ket{\downarrow_z}\rightarrow\ket{\uparrow_z}$} transition at the bias field $B_1$, to produce an initial x-polarized N-atom state $\ket{\psi_0}=e^{-i\frac{\pi}{2} S_y}\ket{\psi_{z0}}=\ket{\!\!\uparrow_{x1}\uparrow_{x2}...\uparrow_{xN}}$. The system evolves for a time $\tau=200$ ms at the initial bias magnetic field $B_1=528.53$ G. Then, a resonant radio-frequency pulse $(\phi)_x$, shifted in phase from the first pulse by $\pi/2$,  rotates the N-atom state about the x-axis~\cite{axis} by a chosen angle $\phi$.  Immediately following this rotation, we reverse the sign of the Hamiltonian by applying a $(\pi)_y$ pulse and tuning the bias magnetic field to a value $B_2=525.83$ G, where the scattering length $a_2=-a$, i.e., $e^{i\pi S_y}H(-a)\,e^{-i\pi S_y}=-H(a)$, from Eq.~\ref{eq:1.1}. After the system  evolves for an additional time $\tau$, the bias field is ramped back to $B_1$, and a final $(\pi/2)_y$ pulse is applied~\cite{Supplement}.    The final state of the N-atom system after the pulse sequence of  Fig.~\ref{fig:protocol}(a) can be written as
\begin{equation}
\ket{\psi_f}=e^{-i\frac{3\pi}{2}S_y}W_\phi(\tau)\ket{\psi_0},
\label{eq:1.2}
\end{equation}
where the $W$-operator  is defined by
\begin{equation}
W_\phi(\tau)=e^{iH(a)\tau}e^{-i\phi\,S_x}e^{-iH(a)\tau},
\label{eq:2.9}
\end{equation}
with $S_x=\sum_{i,\alpha_i}\!s_{x\alpha_i}$ the x-component of the {\it total} spin vector for the $N$-atom sample and $\ket{\psi_0}$ the fully x-polarized state.
After the pulse sequence, the spin densities $n_{\uparrow z}(x)$ and $n_{\downarrow z}(x)$  are measured for a single cloud using two resonant absorption images, separated in time by $10\,\mu$s. We define one repetition of this experimental sequence as a ``single-shot," in Fig.~\ref{fig:protocol}(b)~and~(c). Inverse-Abel transformation of $[n_{\uparrow z}(x)-n_{\downarrow z}(x)]/2$ then measures $S_z(E_i,\phi)\equiv s_{zi}$, for a single shot,  Fig.~\ref{fig:protocol}(c).

 Now we connect the measured $s_{zi}$ to information scrambling~\cite{Schleier-Smith2017,ReyNatPhys2017,ReyNat2019}. Consider a {\it single spin} labelled by $\alpha_i$,  with spin components $s_{x\alpha_i},s_{y\alpha_i},s_{z\alpha_i}$, interacting with the many-body system.  It is straightforward to show~\cite{Supplement},
 \begin{equation}
 C_{\alpha_i}\equiv\bra{\psi_0}[W_\phi(\tau),s_{x\alpha_i}]|^2\ket{\psi_0}=\frac{1}{2}-\bra{\psi_f}s_{z\alpha_i}\ket{\psi_f}.
  \label{eq:commutator}
 \end{equation}
 As the many-body operator $W_\phi$ and the single spin operator $s_{x\alpha_i}$ initially commute, i.e., $[W_\phi(0),s_{x\alpha_i}]=0$, a measurement of $\bra{\psi_f}s_{z\alpha_i}\ket{\psi_f}$ determines how two initially commuting operators fail to commute at a later time, providing a measure of scrambling.

 In the experiments, we measure the {\it collective} spin operators $s_{zi}=\sum_{\alpha_i}s_{z\alpha_i}$, where $\alpha_i\equiv (n_i,n_y,n_z)$ for fixed $n_i$. The corresponding mean square commutator, averaged over the $N_s$ spins with x-energy $E_i$, is~\cite{Supplement}
 \begin{equation}
\hspace*{-0.1in}\frac{1}{N_s}\!\sum_{\alpha_i}C_{\alpha_i}(\phi,\tau)=
\frac{1}{2}\!-\!\frac{1}{N_s}\!\sum_{\alpha_i}\!\bra{\psi_f}s_{z \alpha_i}\ket{\psi_f}.
\label{eq:20.1}
\end{equation}
Further averaging Eq.~\ref{eq:20.1} over atoms with energies within $\Delta E$ of $E_i\equiv E$, we replace the sum on the righthand side by
$S_z(E)\,\Delta E/[n(E)\,\Delta E]$, yielding the measured quantity
\begin{equation}
{\cal F}(E,\phi)\equiv\frac{1}{2}\,\frac{n_{\uparrow_z}(E,\phi)-n_{\downarrow_z}(E,\phi)}
{n_{\uparrow_z}(E,\phi)+n_{\downarrow_z}(E,\phi)}.
\label{eq:20.1b}
\end{equation}
Here, $n(E)=n_{\uparrow_z}(E,\phi)+n_{\downarrow_z}(E,\phi)$ is independent of $\phi$ and ${\cal F}(E,0)=1/2$.

We can extract information about the many-body coherence from  Eq.~\ref{eq:20.1}, by writing the sum on the right-hand side as $\sum_m e^{im\phi}\,B_m$~\cite{Supplement}.  Non-vanishing coefficients $B_m$ correspond to coherence between states for which the x-component $S_x$ of the total angular momentum differs by $m$~\cite{ReyPRL2018,Supplement}. Since the sum is real, $B_{-m}=B_m^*$, we can expand Eq.~\ref{eq:20.1b} for the measured, energy-selected average in the form
\begin{equation}
{\cal F}(E,\phi)=B_0+\sum_{m\geq 1}\! 2\,|B_m|\cos(m\phi+\varphi_m).
\label{eq:18.6}
\end{equation}
In fitting the data with Eq.~\ref{eq:18.6}, we restrict the range of $m$ to 4. We find that the fits  are not improved by further increase of $m$,  consistent with the limited number of $\phi$ values measured in the experiments.

We  measure spin density profiles $n_{\uparrow_z,\downarrow_z}(x,\phi)$ for a scattering length $a=4.24\,a_0$. The data are averaged over $6$ repetitions for each $\phi$, with the $\phi$ values chosen in random order.  We begin by finding the {\it total} number of atoms in each spin state $N_{\uparrow_z,\downarrow_z}(\phi)=\int dx\,n_{\uparrow_z,\downarrow_z}(x,\phi)$ for the protocol of Fig.~\ref{fig:protocol}(a), to find the total collective spin projection $S_z$  versus rotation angle $\phi$, {\it without} energy restriction.  Fig.~\ref{fig:TotalSz}(a) shows the normalized $S_z$ data $F(\phi)=\frac{1}{2}(N_{\uparrow_z}-N_{\downarrow_z})/(N_{\uparrow_z}+N_{\downarrow_z})$ (blue dots) and the fit of Eq.~\ref{eq:18.6} (red curve), which  determines the magnitude (b) and phase (c) of the average coherence coefficients $B_m$. We note that $F(0)\simeq F(2\pi)< 1/2$, the maximum for ideal conditions. This discrepancy arises from small variations in the phase shift of the final $\pi/2$ pulse, which is applied at a finite detuning as the magnetic field is ramped from $B_2$ back to its original value $B_1$~\cite{Supplement}.

\begin{figure*}[htb]
\begin{center}\
\hspace*{-0.25in}\includegraphics[width=5.5in,trim = {0 0 0 0}, clip]{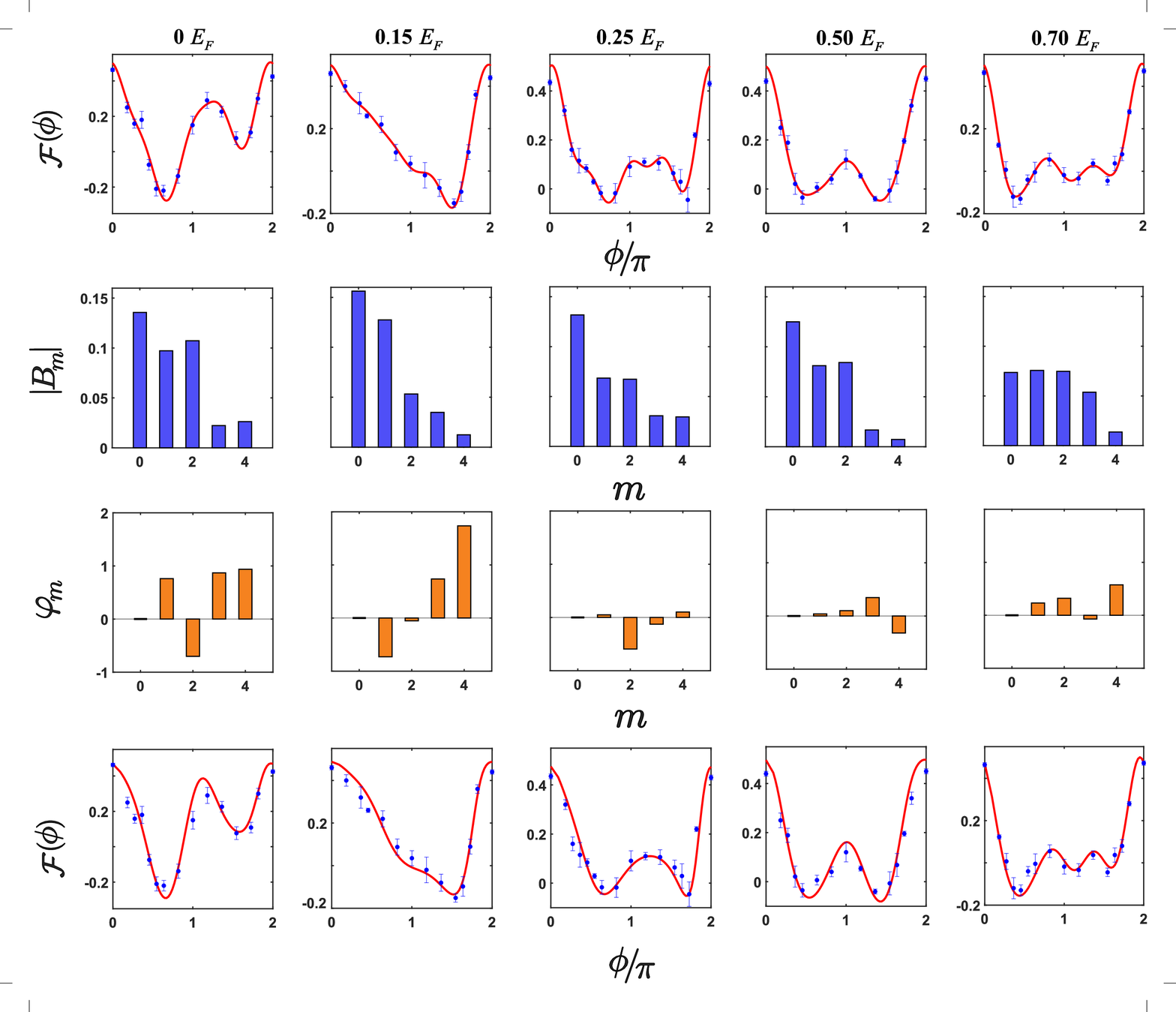}
\end{center}
\caption{Energy-resolved collective spin projection $S_z(E)$ versus rotation angle $\phi$ for spins of selected energies (left to right) $E/E_F=0,0.15,0.25,0.5,0.7$. Here, ${\cal F}(\phi)=\frac{1}{2}[n_\uparrow(E)-n_\downarrow(E)]/[n_\uparrow(E)+n_\downarrow(E)]$. The top row shows the data (blue dots) for a measured scattering length $a=4.24\,a_0$. The red solid curve is the fit of Eq.~\ref{eq:18.6}, which determines the magnitudes of the coherence coefficients $|B_m|$ (second row) and corresponding phases $\varphi_m$ (third row); The bottom row shows the fits (red solid curves) of the mean field model of Ref.~\cite{SaeedPRASpinECorrel} to the data (blue dots), using a scattering length $2.63$ times the measured value and global detunings,  ordered in energy, of $\Delta({\rm Hz}) = 0$, $0.8$, $0.65$, $-0.8$, and $0.15$.
\label{fig:EnergySz}}
\end{figure*}

To check that the measurements are reasonable, we compare the $\phi$-dependent data  of Fig.~\ref{fig:TotalSz} to a fit of our 1D mean field model, which employs a calculated average transverse density $\bar{n}_\perp$ to fit single-pulse spin-wave data with no free parameters~\cite{SaeedPRASpinECorrel}. The model, evaluated with a global detuning $\Delta =0$, is shown in Fig.~\ref{fig:TotalSz}(d). To fit the observed $\phi$ dependence (red solid curve), the model requires a scattering length $a_{eff}\equiv 2.63\,a_{\rm meas}$, i.e., $2.63$ times larger than the measured value $a_{\rm meas}=4.24\,a_0$, which yields the black-dashed curve. The increased $a_{eff}$ may occur because the measured coherence orders with $|m|>1$ arise from interactions,  favoring the largest couplings in a manner that is not predicted by our model.

Fig.~\ref{fig:EnergySz} shows the energy-resolved measurements ${\cal F}(E,\phi)$, obtained by inverse-Abel transformation of the same data. The top row shows significant variation in symmetry and structure as the energy is varied from $E=0$ to $E=0.7\,E_F$. The red solid curves in the first row show the fit of Eq.~\ref{eq:18.6}, which yields the magnitudes of the coherence coefficients $|B_m|$ and the corresponding phases $\varphi_m$.  In the last row, we compare the data to fits of the mean field model~\cite{SaeedPRASpinECorrel}. Again, the model captures the complex $\phi$-dependent shapes of the data with $a_{eff}=2.63\,a_{\rm meas}$, but a different detuning $\Delta$ is needed for each energy. This may be a consequence of averaging data over several detunings $\Delta$, where each $\Delta$ rotates the direction of the $\phi$-rotation axis by $\Delta\tau$~\cite{axis}.

In summary, we have demonstrated a general method for measuring energy-resolved collective spin vectors in an energy-space lattice with effective long-range interactions.  We have shown that an OTOC protocol can be implemented in this system and that many-body coherence can be measured in selected energy-space subsystems. Future measurement of time-dependent correlations between extensive subsets, $C_{ij}(t)\equiv\langle \psi_0|s_{xi}(t)s_{xj}(t)|\psi_0\rangle-\langle\psi_0|s_{xi}(t)|\psi_0\rangle\langle\psi_0| s_{xj}(t)|\psi_0\rangle$, enables a wide variety of protocols, extending correlation measurements in small numbers of trapped ions~\cite{MonroeCorrelProp} to large quantum systems. For an initial x-polarized  product state, $|\psi_0\rangle$, $C_{ij}(t)=0$ for noninteracting systems and for our mean-field model, so that $C_{ij}(t)\neq 0$ signifies beyond mean-field physics. As $C_{ij}(0)=0$,  a scrambling time~\cite{GuoSignaltime,maldacena2016bound} is determined by observing the evolution from the product state to a correlated state.

Primary support for this research is provided by the Air Force Office of Scientific Research (FA9550-16-1-0378) and the National Science Foundation (PHY-2006234). Additional support for the JETlab atom cooling group has been provided by the Physics Division of the Army Research Office (W911NF-14-1-0628) and by the Division of Materials Science and Engineering, the Office of Basic Energy Sciences, Office of Science, U.S. Department of Energy (DE-SC0008646).

$^*$Corresponding author: jethoma7@ncsu.edu


%

\widetext
\setcounter{figure}{0}
\setcounter{equation}{0}
\renewcommand{\thefigure}{S\arabic{figure}}
\renewcommand{\theequation}{S\arabic{equation}}

\appendix
\section{Supplemental Material}

In this supplemental material, we describe the experimental methods for creating and probing an energy-space lattice in a weakly interacting Fermi gas. We demonstrate that out-of-time-order correlation (OTOC) measurements can be implemented in this system and show that many-body spin coherence is measurable, even with imperfect control of the radio-frequency detuning. Finally, we show that an inverse-Abel transform of the measured spin density profiles determines the collective spin vector as a function of energy, enabling many-body coherence measurement in energy-resolved sectors of the energy-space lattice.

\subsection{Experimental Methods}
\label{sec:experiment}

We begin with an optically trapped cloud of $^6$Li atoms in a 50-50 mixture of the two lowest hyperfine states, denoted $\ket{1}$ and $\ket{2}$, which is evaporatively cooled to degeneracy near the $\ket{1}-\ket{2}$ Feshbach resonance at $832.2$ G. To implement the many-body echo protocol, shown in Fig.~1 of the main text, we prepare a degenerate z-polarized initial state, with all atoms in the spin-down hyperfine state $\ket{2}\equiv |\!\downarrow_z\rangle$~\cite{SaeedPRASpinECorrel}. To prepare this state, after evaporative cooling, the magnetic field is ramped to the weakly interacting regime near $1200$ G, and the $\ket{1}\equiv|\uparrow_z\rangle$ spin component is eliminated by means of a resonant optical pulse. Then the bias magnetic field is ramped near $B_0=527.18$ G, where the s-wave scattering length vanishes~\cite{SaeedPRASpinECorrel}.

The s-wave scattering length in Bohr units ($a_0$) has been measured previously~\cite{SaeedPRASpinECorrel},  $a(B)=3.14(8)\,(B-B_0)$, where the bias magnetic field is precisely determined by rf spectroscopy. After preparing initial polarized state, the bias magnetic field is first tuned to $B_1=528.53$ G, where the s-wave scattering length $a = 4.24\,a_0$. We apply a 0.5 ms radiofrequency pulse to rotate the initial spin state by $\pi/2$ about the y-axis, creating an x-polarized state.  After an evolution time $\tau = 200$ ms, we rotate the many-body spin state by an angle $\phi$ about the x-axis, using a radio-frequency (rf) pulse, shifted in phase from the first pulse by $90^o$.  Immediately following this pulse, the Hamiltonian is inverted by applying a $\pi$ rotation about the y-axis and sweeping the bias magnetic field over 5 ms to a value $B_2=525.83$ G, where the scattering length is $- a$. This sweep, over a few gauss, is accomplished using a set of low inductance auxiliary coils, wound concentric with the primary bias field coils. After an additional $\tau = 200$ ms, the bias magnetic field is swept back to its original value $B_1$ over 5 ms and a final $\pi/2$ rotation about the negative y-axis is applied. The density profiles of both spin components $n_{\uparrow_z,\downarrow_z}(x)$ are then immediately measured for a single cloud, using two camera shots separated by $10\,\mu$s. This defines a {\it single-shot} measurement. Subtraction yields the single-shot z-component of the collective spin vector density $S_z(x)=[n_{\uparrow_z}(x)-n_{\downarrow_z}(x)]/2$, which, in the ideal case, corresponds to the x-component $S_x(x)$ just prior to the final $\pi/2$ pulse.

\subsubsection{Trap and Atom Parameters}
\label{sec:parameters}

Our experiments employ a cigar-shaped optical trap with parameters close to those employed in our previous work~\cite{SaeedPRASpinECorrel}, where $\omega_x/2\pi=23$ Hz is the harmonic oscillation frequency along the cigar $x$-axis and $\omega_r/2\pi=625$ Hz is the harmonic oscillation frequency for the transverse directions. The typical total atom number is $N=6.5\times 10^4$  in the spin state $|\!\downarrow_z\rangle$. The global Fermi temperature in the harmonic trap is $T_F=E_F/k_B=\hbar\,(6N\omega_x\omega_r^2)^{1/3}/k_B= 0.73\,\mu$K, with $\sigma_{Fx}=\sqrt{2 E_F/(m\omega_x^2)}=310\,\mu$m the corresponding Fermi radius for the $x$-direction. Fitting the single spin density $n_{\downarrow_z}(x)$ with a finite temperature Thomas-Fermi profile yields $T/T_F=0.32$ for the degenerate sample. For comparison with our mean field model~\cite{SaeedPRASpinECorrel}, it is convenient to approximate the measured $n_{\downarrow_z}(x)$ by a zero-temperature Fermi profile, which yields an effective Fermi radius  of $\sigma_x=345\,\mu$m for the x-direction and a corresponding effective Fermi temperature $T_{Feff}=m \omega_x^2 \sigma_x^2/(2 k_B)=0.90\,\mu$K. The corresponding effective transverse Fermi radius is  $\sigma_r=\sigma_x\,\omega_x/\omega_r=12.7\,\mu$m.

\subsubsection{Radio-Frequency Detuning}

The ideal implementation of the protocol of Fig.~1 of the main text, as described above, assumes a global detuning $\Delta =0$. The $x$ and $y$ axes are defined in a frame rotating at the applied rf frequency, which is stable to 0.1 Hz over several minutes. However, magnetic field drift can change the resonance frequency in between repetitions, causing the Bloch vector to precess in the rotating frame at a rate equal to the global detuning, $\Delta$.
In the experiments, the global detuning $\Delta$ is near resonance at the initial bias magnetic field $B_1$, but changes by several kHz as  auxiliary coils tune the bias magnetic field to $B_2$. 	The resonance frequency shift of the sweep is determined only by the auxiliary coils, hence is independent of the starting field $B_1$.  The net phase shift resulting from the sweep is then controlled by the precise time at which the final $\pi/2$ pulse is applied, which enables the experiments.  In this case, a large, but reproducible phase shift is accumulated during the  evolution time $\tau = 200$ ms at $B_2$ and as the bias field is swept back to $B_1$, just before the final $\pi/2$ pulse is applied.   To adjust the net phase shift, we apply the final $\pi/2$ pulse during the sweep from $B_2$ to $B_1$ as the detuning becomes small $\simeq 100$ Hz, well within the pulse bandwidth, but nonzero. Then, we adjust the time of the final $\pi/2$ pulse by $\simeq 5-10$ ms to select a stable net phase shift near $180^o$ (modulo $2\pi$), where we observe a maximum transfer of atoms from the initially populated state $\ket{2}=\ket{\downarrow_z}$ to the initially unpopulated state $\ket{1}=\ket{\uparrow_z}$ for $\phi = 0$. This is equivalent to a  $-\pi/2$ pulse about the y-axis, rather than a $+\pi/2$ pulse, which we take into account in the data analysis. For single shots with $\phi=0$ or $\phi=2\pi$ in the OTOC protocol, we observe transfer efficiencies from $90$\% to $100$\%, after setting the radio frequency as described below.

To set the radio frequency  close to resonance at the field $B_1$, we initially find the resonance frequency for the transfer of atoms from state $\ket{2}=\ket{\downarrow_z}$ to state $\ket{1}=\ket{\uparrow_z}$ using a single long 50 ms pulse. The observed linewidth is 8 Hz half width at half maximum, enabling an approximate determination of the $\Delta = 0$ frequency within 1 Hz. To keep the rf frequency nominally on resonance as data is collected,  for each choice of $\phi$, we consistently check that the $\phi = 0$ configuration produces maximum transfer of atoms from state $\ket{2}$ to state $\ket{1}$ at the end of the 400 ms total sequence. If not, the rf frequency is changed to compensate for magnetic field drift, which changes the resonance frequency by $\simeq 3.6$ Hz/mG. However, it is not possible to control the detuning at the Hz or sub-Hz level.

Fortunately, drifts in the radiofrequency detuning $\Delta$ are partially mitigated by the $\pi$ pulse at the center of the protocol of Fig.~1 of the main paper, which reverses the net accumulated phase at time $\tau$ for a fixed detuning. If the detuning is stable over the 400 ms duration of the sequence, this accumulated phase is cancelled. Further, we compensate for the phase shift arising from the magnetic field sweep between $B_1$ and $B_2$, as discussed above. In \S~\ref{sec:MBCM}, we discuss the remaining effect of imperfect control of the detuning on the $\phi$-rotation axis.

\subsection{Many-Body Coherence Measurement}
\label{sec:MBCM}
In the following, we discuss the out-of-time-order (OTOC) protocol. We show that imperfect control of the global detuning $\Delta$ over several repetitions of the protocol is equivalent to averaging the {\it direction} of the axis for the $\phi$ rotation. We find that the resulting axis-averaged coherence coefficients are still measurable from the $\phi$-dependent spin density. OTOC protocols are applicable to systems described by a general density  operator. Here, we specialize to the case for a pure initial state $|\psi_0\rangle$, as used in the experiments, where the utility of the OTOC method can be simply understood~\cite{Schleier-Smith2017,ReyNatPhys2017,ReyNat2019}.

Let $W$ and $V$ be two, generally time-dependent, but not necessarily unitary, operators. Consider the two states $|\psi_1\rangle=WV|\psi_0\rangle$ and $|\psi_2\rangle=VW|\psi_0\rangle$, where the operators are applied in reverse order. We define the overlap
\begin{equation}
{\cal F}\equiv\langle\psi_2|\psi_1\rangle=\langle\psi_0|W^\dagger V^\dagger WV|\psi_0\rangle,
\label{eq:9.2}
\end{equation}
which is of the OTOC form~\cite{TimeOrder} . Now consider ${\cal F}+{\cal F}^*=2\,{\cal R}e\{{\cal F}\}$,
\begin{equation}
2\,{\cal R}e\{{\cal F}\}=\langle\psi_0|W^\dagger V^\dagger WV|\psi_0\rangle+\langle\psi_0|V^\dagger W^\dagger VW|\psi_0\rangle.
\label{eq:9.4}
\end{equation}
With $W^\dagger V^\dagger=[W^\dagger, V^\dagger]+V^\dagger W^\dagger$ in the first term and similarly $V^\dagger W^\dagger=-[W^\dagger, V^\dagger]+W^\dagger V^\dagger$ in the second term, we have
$$ 2\,{\cal R}e\{{\cal F}\}=\langle\psi_0|[W^\dagger, V^\dagger] [W,V]|\psi_0\rangle+\langle\psi_0|V^\dagger W^\dagger WV|\psi_0\rangle+\langle\psi_0|W^\dagger V^\dagger VW|\psi_0\rangle.$$
Then, with $[W^\dagger, V^\dagger]=-[W, V]^\dagger$ and $[W,V]^\dagger[W,V]\equiv|[W,V]|^2$, we find generally,
\begin{equation}
2\,{\cal R}e\{{\cal F}\}=\langle\psi_0|V^\dagger W^\dagger WV|\psi_0\rangle+\langle\psi_0|W^\dagger V^\dagger VW|\psi_0\rangle-\langle\psi_0| \,|[W,V]|^2|\psi_0\rangle.
\label{eq:10.4}
\end{equation}

For the special case of unitary operators, $W^\dagger W=1$ and $V^\dagger V=1$,  Eq.~\ref{eq:10.4} takes the simple form
\begin{equation}
{\cal R}e\{{\cal F}\}={\cal R}e\{\langle\psi_0|W^\dagger V^\dagger WV|\psi_0\rangle\}=1-\frac{1}{2}\,\langle\psi_0|\,|[W,V]|^2|\psi_0\rangle.
\label{eq:10.6}
\end{equation}
We note that for unitary operators, the overlapped states are normalized, i.e., $\langle\psi_1|\psi_1\rangle =1$, and $\langle\psi_2|\psi_2\rangle =1$.
Eq.~\ref{eq:10.6} is particularly useful when the left hand side  ${\cal R}e\{{\cal F}(t)\}$ is measurable and time dependent. Then, if $[W,V]=0$ at time $t=0$, a nonzero value for $1-{\cal R}e\{{\cal F}(t)\}$  describes how two initially commuting operators $W$ and $V$ fail to commute at a later time in a many-body system, providing a measure of information scrambling~\cite{Schleier-Smith2017,ReyNatPhys2017,ReyNat2019}.

In the experiments, we determine an average of ${\cal F}$ in Eq.~\ref{eq:9.2}, by measuring the sum of $z$-components of the spin for a selected group of atoms. Consider first  the z-component of the spin for a {\it single} atom,  defined as $s_{z\alpha_i}$. Here, as described in the main text, $\alpha_i$ comprises the 3D vibrational quantum numbers ($n_i,n_y,n_z$) of a {\it single} state, with  $n_i+1/2=E_i/h\nu_x$ specified for the $x$-direction. The final state of the N-atom system, after the pulse sequence of  Fig.~1 (a) of the main text, is
\begin{equation}
\ket{\psi_f}=e^{-i\frac{3\pi}{2}S_y}W_\phi(\tau)\ket{\psi_0},
\label{eq:1.2}
\end{equation}
where $\ket{\psi_0}\equiv e^{-i\frac{\pi}{2} S_y}\ket{\psi_{z0}}=\ket{\!\!\uparrow_{x1}\uparrow_{x2}...\uparrow_{xN}}$ is a pure $x$-polarized $N$-atom state prepared from the initial $z$-polarized state $\ket{\psi_{z0}}$ by the first $\pi/2$ pulse in the OTOC protocol.
Then,
\begin{equation}
\langle\psi_f|s_{z\alpha_i}|\psi_f\rangle=\langle\psi_0|W^\dagger_\phi(\tau)\,s_{x\alpha_i} W_\phi(\tau)|\psi_0\rangle,
\label{eq:ztox}
\end{equation}
where we have used Eq.~\ref{eq:1.2} with $e^{i\frac{3\pi}{2}S_y} s_{z\alpha_i}\,e^{-i\frac{3\pi}{2}S_y}=s_{x\alpha_i}$.
As described in the main text, the unitary $W$-operator in Eq.~\ref{eq:ztox}  is given by
\begin{equation}
W_\phi(\tau)=e^{iH(a)\tau}e^{-i\phi\,S_x}e^{-iH(a)\tau},
\label{eq:2.9}
\end{equation}
where $H(a)$ is the Hamiltonian and  $S_x=\sum_{j,\alpha_j}\!s_{x\alpha_j}$ is the x-component of the {\it total} spin vector for the $N$-atom sample.
Physically, $W$ rotates the entire $N$-atom system by an angle $\phi$ about the $x$-axis, in between forward and backward time evolutions of duration $\tau$.

Now we can show that right hand side of Eq.~\ref{eq:ztox} is of the OTOC form, by using Eq.~\ref{eq:9.2} with $V=\sigma_{x\alpha_i}$,  where $\sigma_{x\alpha_i}=2\,s_{x\alpha_i}$ is the Pauli matrix for a {\it single} spin labeled by $\alpha_i$, and $s_{x\alpha_i}$ is the corresponding spin operator.  Then,  $V=V^\dagger$ is hermitian and unitary, $V^\dagger V=1$,  for each spin.  For the $x$-polarized state $|\psi_0\rangle$, $\sigma_{x\alpha_i}|\psi_0\rangle=V|\psi_0\rangle=|\psi_0\rangle$ for each $\alpha_i$ and ${\cal F}\rightarrow\langle\psi_0|W^\dagger V W|\psi_0\rangle$ is real. With $V=2\,s_{x\alpha_i}$, Eq.~\ref{eq:10.6} for each spin yields
\begin{equation}
{\cal F}_{\alpha_i}(\phi,\tau)=2\langle\psi_0|W^\dagger_\phi(\tau)s_{x\alpha_i} W_\phi(\tau)|\psi_0\rangle=1-\frac{4}{2}\,\langle\psi_0| \,|[W_\phi(\tau),s_{x\alpha_i}]|^2|\psi_0\rangle.
\label{eq:12.6}
\end{equation}
Eq.~\ref{eq:ztox} and  Eq.~\ref{eq:12.6} then give the mean square commutator for a single spin in terms of the measured $z$-component,
\begin{equation}
C_{\alpha_i}(\phi,\tau)=\bra{\psi_0}\,|[W_\phi(\tau),s_{x\alpha_i}]|^2\ket{\psi_0}=\frac{1}{2}-\bra{\psi_f}s_{z\alpha_i}\ket{\psi_f}.
\label{eq:commutator}
\end{equation}

\subsubsection{Detuning dependence of the Coherence Coefficients}

We can understand the effect of finite global detuning $\Delta$ on the coherence coefficients by considering the measurement of the z-projection of a single spin $s_{z\alpha_i}$, where $\alpha_i$ denotes the 3D vibrational state of an atom with axial energy $E_{xi}$, as discussed above. After the pulse sequence, Eq.~\ref{eq:ztox} and Eq.~\ref{eq:12.6} show that
\begin{equation}
{\cal F}_{\alpha_i}(\phi)\equiv 2\bra{\psi_f}s_{z\alpha i}\ket{\psi_f}=2\,\bra{\psi_0}W^\dagger_\phi(\tau)\,s_{x\alpha_i}W_\phi(\tau)\ket{\psi_0}.
\label{eq:Fi}
\end{equation}

To explicitly display the detuning dependence of the measurement, we write the Hamiltonian in Eq.~1 of the main text as $H(a,\Delta)\equiv H(a,0)-\Delta\,S_z$, where $S_z=\sum_{j,\alpha_j}\!\hat{s}_{z\alpha_j}$ is the z-component of the total spin vector. Then, since $[H(a,0),S_z]=0$, we have from Eq.~\ref{eq:2.9}
\begin{equation}
W_\phi(\tau)=e^{iH(a,0)\tau}\,e^{-i\Delta\tau\,S_z}e^{-i\phi\,S_x}e^{i\Delta\tau\,S_z}e^{-iH(a,0)\tau}=
e^{iH(a,0)\tau}e^{-i\phi\,S_{x'}}e^{-iH(a,0)\tau}.
\label{eq:2.9b}
\end{equation}
Here, the phase shift $\Delta\tau$ is accumulated during the time $\tau$ between the first $\pi/2$ pulse and the $\phi$ rotation. We see that a nonzero detuning changes the axis for the $\phi$ rotation from $x$ to $x'$, with $S_{x'}\equiv S_x\,\cos(\Delta\tau)+S_y\,\sin(\Delta\tau)$.

From the structure of $W_\phi(\tau)$ in Eq.~\ref{eq:2.9b}, we see that for each detuning $\Delta$, we can expand Eq.~\ref{eq:Fi} using matrix elements  in a total angular momentum eigenstate basis $|J,M\rangle_{x'}$, with $S_{x'}|J,M\rangle_{x'}=M\,|J,M\rangle_{x'}$, where we suppress all other quantum numbers that define the states, such as intermediate angular momenta. Then,  Eq.~\ref{eq:Fi} for a single spin in state $\alpha_i$ can be written as
\begin{equation}
{\cal F}_{\alpha_i}(\phi)=2\, \sum_{m} B^{(\alpha_i)}_m\,e^{i m\,\phi},
\label{eq:18.6S}
\end{equation}
where the integer $m=M'-M$ is the difference of the total angular momentum projections along the $x'$ axis, and
\begin{equation}
B^{(\alpha_i)}_m =\sum_{J,J',M}\,_{x'}\langle J M|\rho(\tau)|J'M+m\rangle_{x'}\,_{x'}\langle J' M+m|s_{x\alpha_i}(\tau)|JM\rangle_{x'}.
\label{eq:Bmi}
\end{equation}
Here, $\rho(\tau)=e^{-iH(a,0)\tau}\ket{\psi_0}\bra{\psi_0}e^{iH(a,0)\tau}$ is the density operator at time $\tau$ and
$s_{x\alpha_i}(\tau)\equiv e^{-iH(a,0)\tau}\,s_{x\alpha_i}\,e^{iH(a,0)\tau}$. For $\phi=0$, using the completeness of the total angular momentum states, we have $2\,\sum_{m} B^{(\alpha_i)}_m=2\,\bra{\psi_0}s_{x\alpha_i}\ket{\psi_0}=1$, i.e., $\sum_{m} B^{(\alpha_i)}_m=\frac{1}{2}$, as in the main text. Further, $B^{(\alpha_i)}_{-m}=B^{(\alpha_i)*}_{m}$, as required for real ${\cal F}_{\alpha_i}(\phi)$.

Without interactions, $a=0$, the Hamiltonian reduces to an energy-dependent rotation about the $z-$axis. As $s_{x\alpha_i}$ is a rank one operator, for $a=0$, $m=0,\pm 1$ only, corresponding to the $\phi$-dependent projection of each spin along the $x$-axis. However, for the interacting system, $a\neq 0$, collisions create coherence between spins with different energies and nonvanishing density matrix elements between total angular momentum states with $|m|=|M'-M|>1$.

In the experiments, we measure the sum of Eq.~\ref{eq:18.6S} over the $N_s$ atoms in different occupied transverse modes $n_y,n_z$ with a fixed x-energy $E_i$. For each $E_i$, we have an average coefficient
\begin{equation}
B_m=\frac{1}{N_s}\sum_{\alpha_i} B^{(\alpha_i)}_m.
\label{eq:BmSum}
\end{equation}
Then, averaging Eq.~\ref{eq:BmSum} over atoms with energies $E$ within $\Delta E$ of $E_i$, we obtain the same form as Eq.~8 of the main text,
\begin{equation}
{\cal F}(\phi)=\sum_{m} B_m\,e^{i m\,\phi}=B_0+\sum_{m\geq 1} 2\,|B_m|\,\cos(m\phi+\varphi_m),
\label{eq:average}
\end{equation}
where ${\cal F}(\phi=0)=\frac{1}{2}$.
For an average of several shots with varying detunings, as utilized in the experiments to measure the $\phi$ dependence of the spin density, the expansion coefficients $B^{(\alpha_i)}_m$ of Eq.~\ref{eq:Bmi} are simply averaged over a range of rotation axes $x'$ to obtain the axis-averaged $B_m$ in Eq.~\ref{eq:average}.  This axis averaging, and the sum over a small range of spin energies near $E$ in Eq.~\ref{eq:BmSum}, does not change the general $\phi$-dependent structure of Eq.~\ref{eq:18.6S}, which enables measurements of the average coherence coefficients for imperfectly controlled detuning $\Delta$, as shown in the main text.

We find that the spatial profiles predicted by our mean-field model, see Fig.~\ref{fig:Allprofiles}, are sensitive to the detuning $\Delta/2\pi$ at the fraction of 1 Hz level for $\tau=200$ ms,  where $\Delta\tau$ is determined modulo $2\pi$ in the OTOC experiments. For ensemble-averages of a large number of repetitions of the experiments, the mean field model then can be used as a marker to constrain the detuning in subsets of the data. This was not done in the present experiments, where only 6 repetitions were taken for each $\phi$.

\subsection{Inverse-Abel Transform Method}

The energy-dependent collective spin vector ${\mathbf{S}}(E)$  is determined as the inverse-Abel transform of the measured spatial profile of the spin density ${\mathbf{S}}(x)$. In using this method, we assume that the measured axial spin density  profiles $n_\sigma(x)$, with $\sigma=\uparrow_z,\downarrow_z$, are given in the continuum limit by,
\begin{equation}
n_\sigma(x)=\int dE\,|\phi_E(x)|^2\,n_\sigma(E).
\label{eq:spindensity}
\end{equation}
Here, we have defined $E_x\equiv E$ to simplify the notation, and $N_\sigma=\int dx\,n_\sigma(x)=\int dE\,n_\sigma(E)$ is the total number  in each spin state $\sigma$. For our experimental parameters, where $E_F/\hbar\omega_x\simeq 650$, the continuum approximation is justified.

In addition to the continuum approximation, an important feature of Eq.~\ref{eq:spindensity} is the assumption that there is {\it neglible coherence} between single-atom {\it energy} states. This is justified, in part, by the energy-conserving regime of a very weakly interacting Fermi gas. Physically, each atom remains on its respective energy ``site," $E_i$. As shown in Ref.~\cite{SaeedPRASpinECorrel} and in Fig.~\ref{fig:Allprofiles} below, the sum of the spatial profiles for the two spin densities is time-independent and thermal, despite the small scale spatial structure observed in the spin density $S_z(x)$. Further, this assumption yields predictions in very good agreement the measured spin density for single pulse experiments, using a mean field model for ${\mathbf{S}}(E,t)$ and a WKB approximation for $|\phi_E(x)|^2$~\cite{SaeedPRASpinECorrel}.

\subsubsection{Two-Body Collision Matrix Elements}

We further substantiate that energy-space coherence is negligible by estimating the matrix elements between energy states for a scattering interaction between a spin-up atom and a spin-down atom. In a harmonic trap, the unperturbed two-atom Hamiltonian separates into center of mass and relative motion parts. Then, the center of mass energy state does not change in a two-body collision, as the scattering interaction depends only on the relative coordinate, $\mathbf{r}$. Hence, we estimate the coupling between energy states from the matrix element of the s-wave contact interaction $H'\equiv 4\pi\hbar^2 a/m\,\delta(\mathbf{r})$  between states of relative motion $\psi_{k,m_r,l}(x,r_\perp,\varphi)$ in a cylindrically symmetric harmonic potential. Here,  the vibrational quantum numbers are $k$ for the $x$ coordinate in the weakly confined direction and $m_r$ for the transverse $r_\perp$ coordinate in the tightly confined directions. For the matrix element to be nonzero, we require $\psi_{k,m_r,l}(0)\neq0$. This requires $k=2n$ {\it even} and the azimuthal angular momentum $l=0$. Suppressing $l=0$, the relevant states for reduced mass $\mu=m/2$ are
\begin{equation}
\psi_{2n,m_r}(x,r_\perp)=\phi_{2n}(x)\chi_{m_r}(r_\perp)\equiv\phi_{2n}(z)\,\frac{e^{-\frac{r_\perp^2}{4 l_\perp^2}}}{\sqrt{2\pi l_\perp^2}}\,L^0_{m_r}\!\!\left(\frac{r_\perp^2}{2\, l_\perp^2}\right),
\label{eq:11.6}
\end{equation}
where $l_\perp=\sqrt{\hbar/(m\,\omega_r)}$ is the single-atom harmonic oscillator length for the tightly confined directions
and\\ $\int_0^\infty 2\pi\,r_{\!\perp} dr{\!_\perp}\,\chi^*_{m_r'}(r_\perp)\chi_{m_r}(r_\perp)=\delta_{m_r',m_r}$ for transverse $l=0$ states of energy $E_{m_r}^0=(2 m_r+1)\,h\nu_r$.
The $\phi_{2n}(x)$ are normalized axial harmonic oscillator states with $|\phi_0(0)|^2=1/(l_x\sqrt{2\pi})$
and $l_x=\sqrt{\hbar/(m\,\omega_x)}$. As the Laguerre polynomial for $l=0$ is
\begin{equation}
L_{m_r}^0(\rho)=\sum_{k=0}^{m_r}\frac{(-\rho)^k\,m_r!}{(k!)^2(m_r-k)!},
\end{equation}
we have $L_{m_r}^0(0)=1$, independent of $m_r$.
Then, suppressing $l=0$,
\begin{equation}
\langle 2n',m_r'|H'|2n,m_r\rangle=\frac{4\pi\hbar^2\,a}{m}\frac{g_n'g_n}{(2\pi)^{3/2}\,l_\perp^2\,l_x}\simeq \hbar\omega_x\sqrt{\frac{2}{\pi}}
\,\frac{\omega_r}{\omega_x}\,\frac{a}{l_x}\,\frac{1}{\sqrt{\pi}}\,\frac{1}{(n'\,n)^{1/4}}.
\end{equation}
Here $|g_n|^2=|\phi_{2n}(0)|^2/|\phi_0(0)|^2 =(2n)!/[2^{2n}(n!)^2)]$. The result on the right is obtained for large $n$, where the Stirling approximation $j!=\sqrt{2\pi}\,j^{j+1/2}e^{-j}(1+\frac{1}{12j})$ yields  $|g_n|^2=1/\sqrt{\pi\,n}$. Note that the matrix element is independent of the transverse vibrational quantum numbers $m_r$ and $m_r'$. For our experimental parameters, the matrix element is greatly suppressed, as $l_x= 8.5\,\mu$m. For $a=4.24\,a_0$, $a/l_x\simeq 2.6\times 10^{-5}$. Then, with $\omega_r/\omega_x=27$, $\langle 2n',m_r'|H'|2n,m_r\rangle/h\nu_x\simeq 3.2\times 10^{-4}/(n'\,n)^{1/4}$. For typical $n'\simeq n=100$, $\langle 2n',m_r'|H'|2n,m_r\rangle\rightarrow 3.2\times 10^{-5}\,\hbar\omega_x$. Hence, the coupling between single particle energy states is negligible compared to the axial energy scale $\hbar\omega_x$, where $\omega_x/2\pi=23$ Hz, so that we do not expect coherence between single-particle energy states $\phi_E(x)$.

\begin{figure}[htb]
\begin{center}\
\hspace*{-0.1in}\includegraphics[width=6.0in]{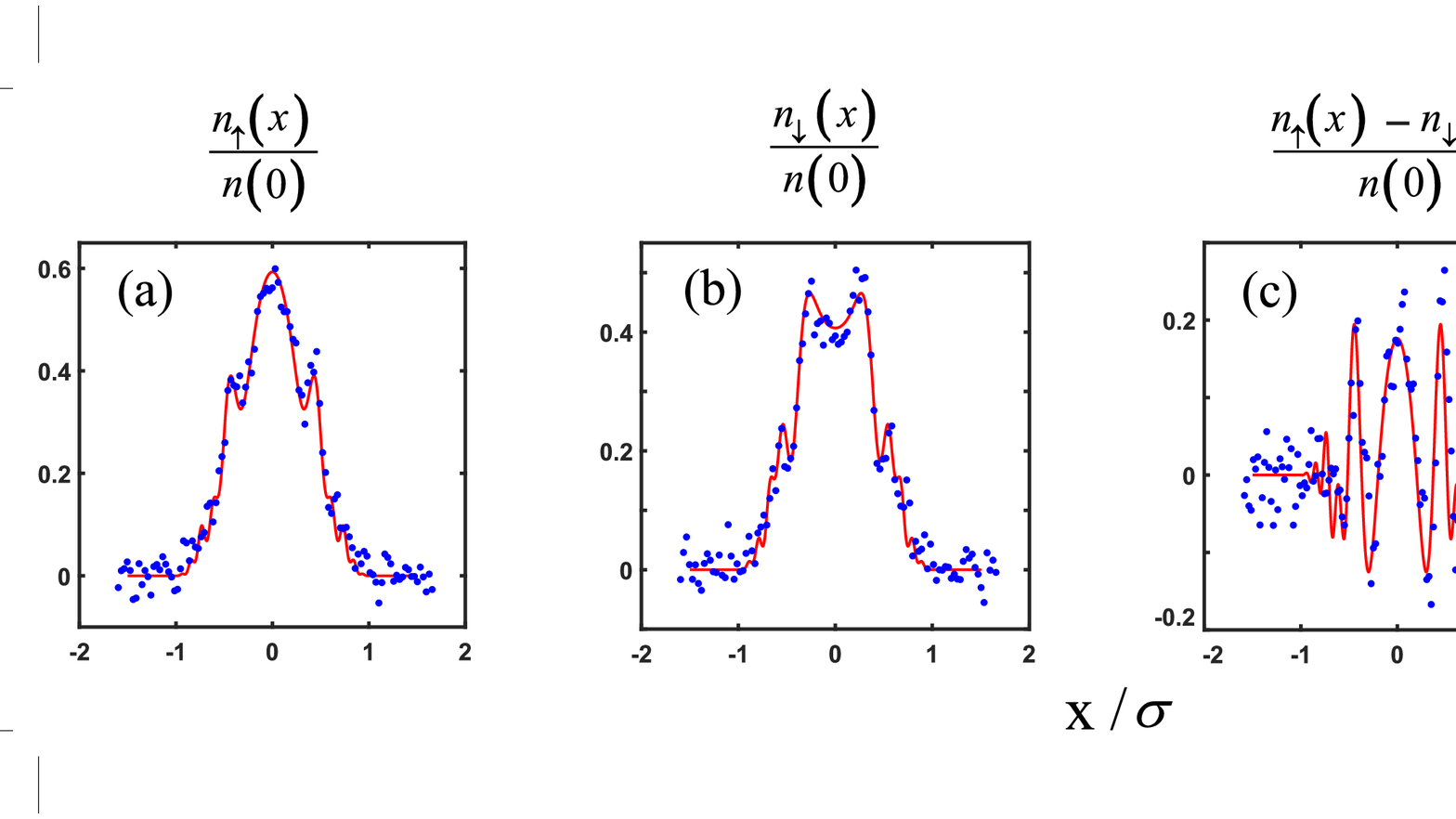}
\end{center}
\caption{Spin density profiles measured for a single shot  with $a=4.24\,a_0$ and $\phi=\pi$ (blue dots) in units of the total central density $n(0)$. (a) $n_\uparrow(x,\phi=\pi)$; (b) $n_\downarrow(x,\phi=\pi)$; (c) Difference of the density profiles $[n_\uparrow(x,\phi=\pi)-n_\downarrow(x,\phi=\pi)]/n(0)$; (d) Total density $n(x)=n_\uparrow(x,\pi)+n_\downarrow(x,\pi)$ in units of the central density $n(0)$. Despite the complex spatial structure in the individual spin density profiles, the total density remains thermal.  The red curves show the predictions of the mean field model of Ref.~\cite{SaeedPRASpinECorrel} using a scattering length 2.35 times the measured value of $4.24\,a_0$ and a global detuning $\Delta/2\pi= 0.27$ Hz.
\label{fig:Allprofiles}}
\end{figure}

To illustrate these ideas,  Fig.~\ref{fig:Allprofiles} shows the single-shot spin density profiles taken after the full OTOC pulse sequence of Fig.~1 of the main paper with $a=4.24\,a_0$ and $\phi=\pi$ (blue dots). Despite the complex structure observed in the spatial profiles for the individual spin densities, which arises from spin coherence,  the total density, shown on the right hand side, remains in a thermal distribution, consistent with the assumption of no energy-space coherence.

\subsubsection{WKB Approximation}

In the continuum limit, where the harmonic oscillator energy level spacing is small compared to the energy scale, as discussed above, the harmonic oscillator wave functions can be evaluated using a WKB approximation. Neglecting the rapid spatial oscillation arising from the WKB phase for large  $n_x\simeq E/\hbar\omega_x$, the normalized probability densities are~\cite{SaeedPRASpinECorrel},
\begin{equation}
|\phi_E(x)|^2=\frac{\omega_x}{\pi}\int_0^\infty dp_x\,\delta\left(E-\frac{p_x^2}{2m}-\frac{m\omega_x^2}{2}x^2\right).
\label{eq:3.2}
\end{equation}
Then the spin densities of Eq.~\ref {eq:spindensity} take the quasi-classical form
\begin{equation}
n_\sigma(x)=\frac{\omega_x}{\pi}\int_0^\infty dp_x\,n_\sigma\left(\frac{p_x^2}{2m}+\frac{m\omega_x^2}{2}x^2\right),
\label{eq:spindensity1}
\end{equation}
which is an Abel-transform of $n_\sigma(E)$, i.e.,  the y-integral of a function of $r^2=x^2+y^2$. Hence, an inverse-Abel transform  determines the energy-dependent collective spin component $n_\sigma(E)$ from the measured spatial profile $n_\sigma(x)$.

Eq.~\ref{eq:spindensity1} is equivalent to the local density approximation for the spatial profile. For example, consider the normalized $T=0$ one-dimensional energy $E=E_x$ distribution for a single spin component Fermi gas, which is obtained from the three-dimensional $T=0$ energy distribution by integrating over $E_y$ and $E_z$,
\begin{equation}
n_\uparrow(E)=N_\uparrow\,\frac{3}{E_F}\left(1-\frac{E}{E_F}\right)^2\theta\left(1-\frac{E}{E_F}\right).
\label{eq:5.2}
\end{equation}
Inserting Eq.~\ref{eq:5.2} into Eq.~\ref{eq:spindensity1} and carrying out the momentum $p_x$ integral, we easily obtain
\begin{equation}
n_\uparrow(x)=\frac{16}{5\pi}\frac{N_\uparrow}{\sigma_{Fx}}\left(1-\frac{x^2}{\sigma_{Fx}^2}\right)^{\frac{5}{2}}\,
\theta\left(1-\frac{x^2}{\sigma_{Fx}^2}\right),
\label{eq:6.4}
\end{equation}
where $\sigma_{Fx}=\sqrt{2E_F/m}/\omega_x$, with $\omega_x=2\pi\nu_x$. Eq.~\ref{eq:6.4} is the normalized one dimensional spatial profile for a Fermi gas at $T=0$, which is shown in Fig.~\ref{fig:Allprofiles}(d). Here, the $T=0$ fit determines an effective $T=0$ Fermi radius and corresponding Fermi temperature, as discussed in \S~\ref{sec:parameters}, which we use to simplify the implementation of the mean field model~\cite{SaeedPRASpinECorrel}.

To extract the spin projection $S_z(E)=[n_{\uparrow_z}(E)-n_{\downarrow_z}(E)]/2$ from the data, the measured spatial profile $S_z(x)=[n_{\uparrow_z}(x)-n_{\downarrow_z}(x)]/2$  is first symmetrized by folding about $x=0$ and then an inverse-Abel transform is implemented without employing derivatives by using the method described in Ref.~\cite{NewAbelInversion}. For this method, the unknown energy distribution is expanded in a series of cosine-functions of the form
\begin{equation}
f_n(E)=1-(-1)^n\,\cos(n\, \pi \sqrt{E/E_F}),
\label{eq:cosine}
\end{equation}
with an amplitude $A_n$ for each $f_n(E)$. The $A_n$ are calculated by least-squares-fitting the Abel-transformed series to the measured data, which yields a matrix equation for the $A_n$.

The energy resolution of the inverse-Abel transform method is limited by the signal to background ratio of the spatial profiles and the position resolution of the imaging system, which limits the maximum value of $n$ in the series. For a fit employing $n_{\rm max}$ cosine functions, the resolution $\Delta E$ is estimated from $\Delta E\,\partial_E(n_{\rm max}\pi \sqrt{E/E_F})=1$, which  yields,
\begin{equation}
\frac{\Delta E}{E_F}=\frac{2}{n_{\rm max}\,\pi}\sqrt{\frac{E}{E_F}}.
\label{eq:resolution}
\end{equation}
For $n_{\rm max}=8$, as employed to analyze the data, and the average energy $E=\bar{E}_x=E_F/4$, we find $\Delta E/E_F\simeq 0.04$, which is adequate for resolving the energy-space partitions. We note that the resolution scales as $\sqrt{E/E_F}$, with the useful property of producing a narrow bandwidth near $E=0$, where many radial modes are occupied and $n_\sigma(E)$ is large, and a larger bandwidth for $E\simeq E_F$, where very few radial modes are occupied and $n_\sigma(E)$ is small. A similar resolution limit is obtained from the image spatial resolution $\Delta x =5\,\mu$m (or the data point spacing), by assuming  $\Delta E/E_F\simeq\Delta(m\omega_x^2 x^2/2)/E_F= 2x\,\Delta x/\sigma_x^2$. For our experiments, $x\leq\sigma_x=345\,\mu$m, $\Delta E\leq 0.03$.

\begin{figure}[htb]
\begin{center}\
\hspace*{-0.2in}\includegraphics[height=1.75in]{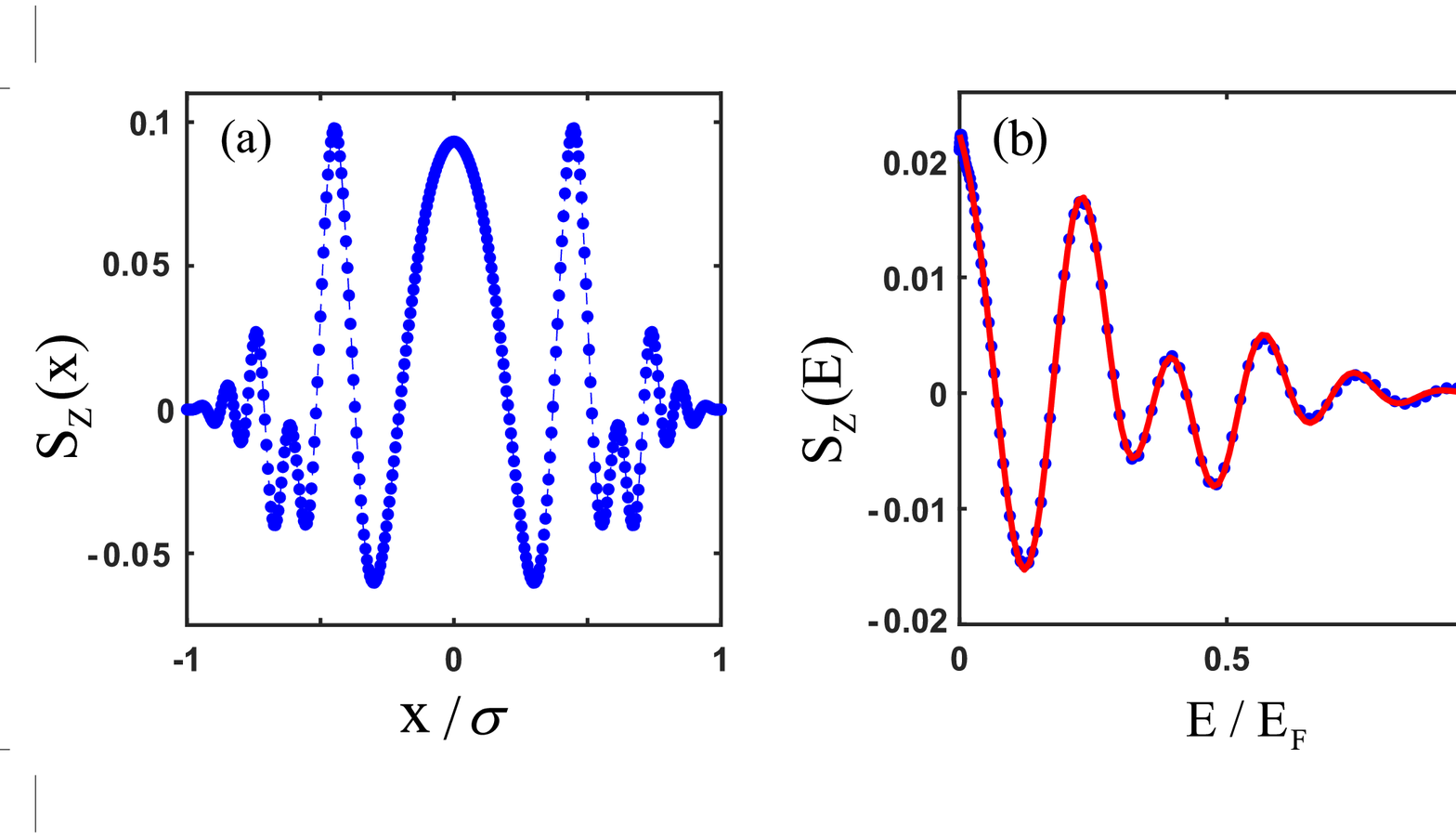}
\end{center}
\caption{Testing the inverse Abel-transform method. Using a mean field model, spin density ``data" (a) for $S_z(x,\phi=\pi)$ are generated for the protocol of Fig.~1 of the main paper, with the same x-spacing as the actual data. Inverse Abel-transformation  yields (b) $S_z(E,\phi=\pi)$  (blue dots), which closely matches the input $S_z(E,\phi=\pi)$ (red curve) from the mean field model that was used to generate the model data for the spin density spatial profile.
\label{fig:comparison}}
\end{figure}

To test the inversion method, we generate model ``data" for $S_z(x,\phi=\pi)$, Fig.~\ref{fig:comparison}(a), with the same $x$ spacing as the real data. Here, $S_z(x,\phi=\pi)$ is determined from $S_z(E,\phi=\pi)$ by analogy to Eq.~\ref{eq:spindensity1}, using the mean field model of Ref.~\cite{SaeedPRASpinECorrel} to predict $S_z(E,\phi=\pi)$ for $\phi =\pi$, scattering length $a=2.35\times 4.24\,a_0$, and global detuning $\Delta/2\pi=0.27$ Hz, as used in the fits of Fig.~\ref{fig:Allprofiles}. Inverting the model data for $S_z(x,\phi=\pi)$, we find the result shown as the blue dots of Fig.~\ref{fig:comparison}(b). For the inversion, we start with a small number of cosine terms and increase the number until the agreement with the exact input  $\mathbf{S}(E,\phi=\pi)$ (red curve) shows no further improvement. Using 20 cosine terms, we find that the $S_z(E,\phi=\pi)$ obtained from the spatial profile by inversion (blue dots) is in close agreement with the exact input from the mean field model (red curve) that was used to generate the spatial profile.

\begin{figure}[htb]
\begin{center}\
\hspace*{-0.2in}\includegraphics[height=1.75in]{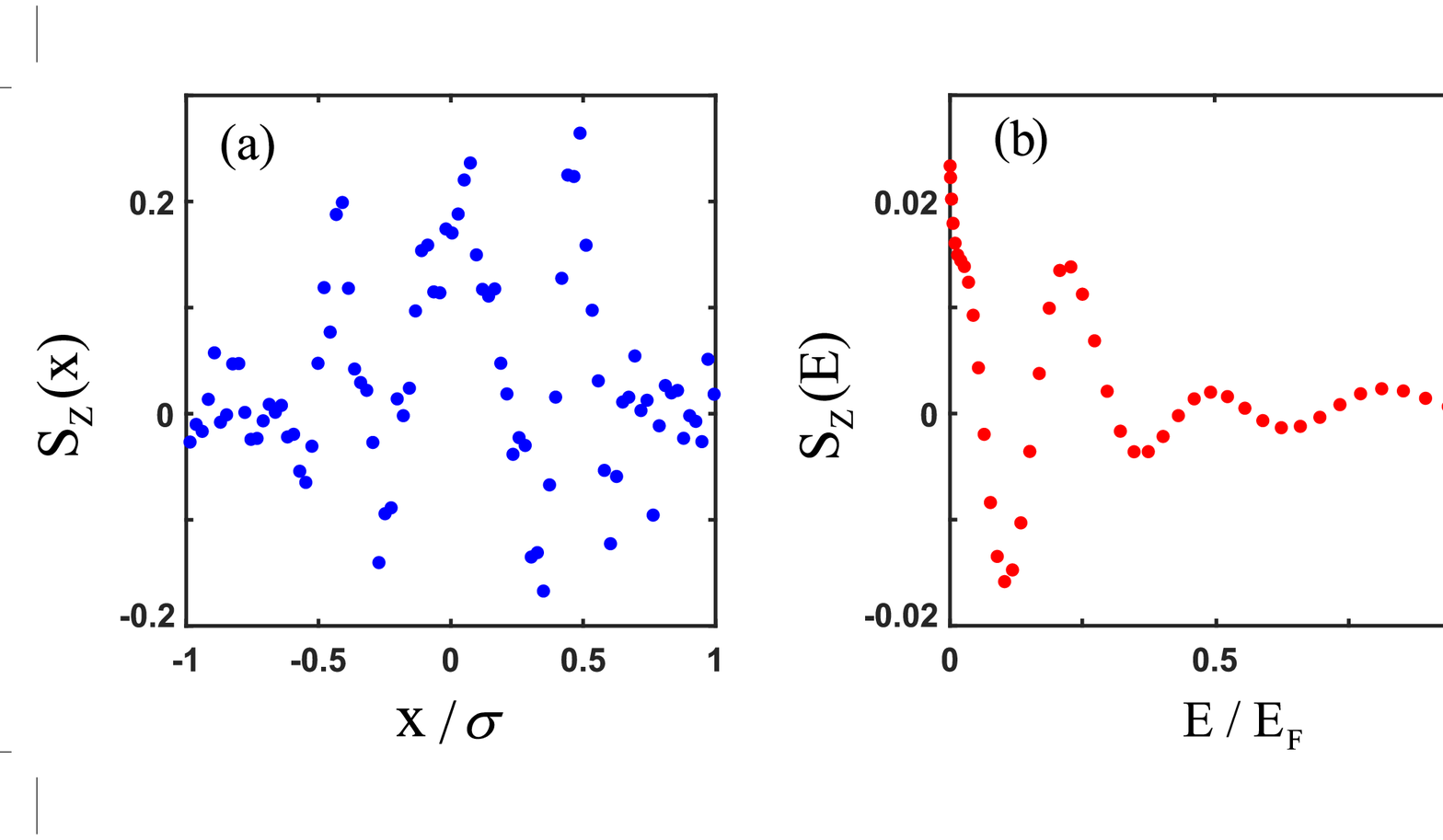}
\end{center}
\caption{Extracting the energy-dependent collective spin component $S_z(E,\phi=\pi)$ for a single shot. (a) Measured single-shot spin density $S_z(x,\phi=\pi)$  for the protocol of Fig.~1 with $a=4.24\,a_0$ and $\phi=\pi$. (b) Inverse Abel-transformation of (a) with $8$ cosine terms yields $S_z(E,\phi=\pi)$ (red dots). (c) $S_z(x,\phi=\pi)$ (red curve) generated from the extracted  $S_z(E,\phi=\pi)$ is consistent with the input spin density data (blue dots).
\label{fig:SZXEData}}
\end{figure}

Next, we apply the Abel-transform method to find the energy-dependent spin component $S_z(E,\phi)$ from the measured spin density $S_z(x,\phi)$ for a single shot, Fig.~\ref{fig:SZXEData}(a). In the data analysis, we employ 8 cosine terms for the inverse-Abel transform of $S_z(x,\phi)$. Further increase in the number of cosine terms is limited by noise in the data and increases the noise in the transform.
 We check the consistency of the extracted 8-term inverse-Abel transform $S_z(E,\phi)$, Fig.~\ref{fig:SZXEData}(b),  by Abel-transformation  to generate the corresponding spatial profile $S_z(x,\phi=\pi)$.   Fig.~\ref{fig:SZXEData}(c) shows that the spatial profile (red curve) generated from the extracted $S_z(E,\phi)$ is consistent with the input spatial profile (blue dots), i.e., the 8 term expansion is adequate to reproduce the small scale spatial structure of the input data.

\begin{figure}[h]
\begin{center}\
\hspace*{-0.2in}\includegraphics[height=2.25in]{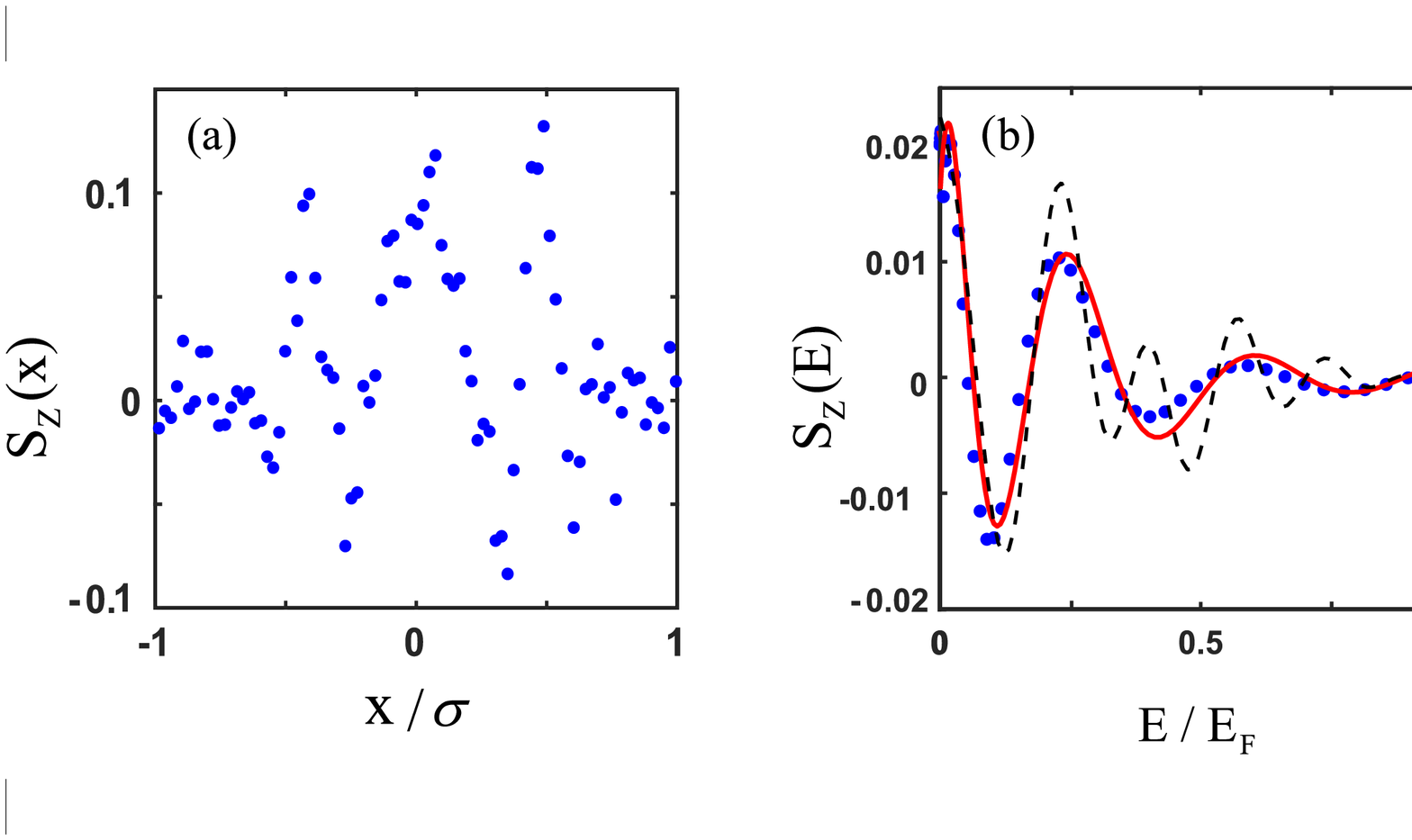}
\end{center}
\caption{Comparison of the extracted energy-dependent collective spin component $S_z(E,\phi=\pi)$ for a single shot with the mean field model. (a) Measured single-shot spin density $S_z(x,\phi=\pi)$  for the protocol of Fig.~1 with $a=4.24\,a_0$ and $\phi=\pi$. (b) Inverse Abel-transformation of (a) with 8 cosine terms yields $S_z(E,\phi=\pi)$ (blue dots). The red curve is the $S_z(E,\phi=\pi)$ obtained with an 8-term inverse-Abel transform of the spatial profile  $S_z(x,\phi=\pi)$ predicted by the  mean field model of Ref.~\cite{SaeedPRASpinECorrel}, for a scattering length $a=2.35$ times measured value $a=4.24\,a_0$ and a global detuning $\Delta/2\pi=0.27$ Hz. The black-dashed curve shows the  $S_z(E,\phi=\pi)$ that is obtained directly from the mean field model, i.e., without inverse-Abel transform of the predicted spatial profile.
\label{fig:SZEDataModel}}
\end{figure}

 Fig.~\ref{fig:SZEDataModel}(b) shows a comparison between the $S_z(E,\phi=\pi)$ curves extracted using an 8-term inverse-Abel transform of the single shot data $S_z(x,\phi=\pi)$ of Fig.~\ref{fig:SZEDataModel}(a) and by an 8-term inverse-Abel transform of the spatial profile predicted by the mean field model. We find that the predicted and measured shapes are in good agreement. However, as noted in Fig.~\ref{fig:Allprofiles}, the mean field model requires a scattering length that is $2.35$ times the measured value.


\begin{thebibliography}{30}
\bibitem [{Num()}]{NumberperSite}%
  \BibitemOpen
  \href@noop {} {}\bibinfo {note} {The number of atoms $\Delta n_i$ at site
  $i$, i.e., in x-mode $E_i$, summed over $y,z$ transverse modes, is $\Delta
  n_i\simeq\frac{3\,\Delta E_i\,N}{E_F}(1-E_i/E_F)^2$. With $\Delta
  E_i=\hbar\omega_x$, $N=6.5\times 10^4$, and $E_i=E_F/2$, we find $\Delta
  n_i\simeq 74$.}\BibitemShut {Stop}%
\bibitem [{gAv()}]{gAv}%
  \BibitemOpen
  \href@noop {} {}\bibinfo {note} {Here, we employ a continuum approximation
  with $\bar{g}\equiv\int\!dnP(n)\!\int\!dn'P(n')g(n,n')$, in the notation of
  Ref.~\cite{SaeedPRASpinECorrel}.}\BibitemShut {Stop}%
\end{thebibliography}
\end{document}